\pdfoutput=1
\documentclass[aps,prl,twocolumn,showpacs,preprintnumbers,amsmath,amssymb]{revtex4}
\usepackage{graphicx,subfigure}
\usepackage{dcolumn}
\usepackage{bm}
\usepackage{ulem}
\usepackage{color}

\def\be{\begin{equation}}
\def\ee{\end{equation}}
\def\ba{\begin{eqnarray}}
\def\ea{\end{eqnarray}}

\newcommand \vect[1]{{\boldsymbol{#1}}}
\begin{document}

\title{The spin excitations of the block-antiferromagnetic K$_{0.8}$Fe$_{1.6}$Se$_2$}
\author{Yi-Zhuang You$^{1,3}$, Hong Yao$^{1,2}$, Dung-Hai Lee$^{1,2}$}
\affiliation{
$^{1}$ Department of Physics, University of California at Berkeley, Berkeley, CA 94720, USA,\\
$^{2}$ Materials Sciences Division, Lawrence Berkeley National Laboratory, Berkeley, CA 94720, USA\\
$^{3}$Institute for Advanced Study, Tsinghua University, Beijing, 100084, China}

\date{\today}
\begin{abstract}
We study the spin excitations of the newly discovered block-antiferromagnetic state in K$_{0.8}$Fe$_{1.6}$Se$_2$ using an effective spin Hamiltonian suggested in the literature. Interestingly in addition to the usual Goldstone mode, there exist three other  ``optical'' spin wave branches. These spin excitations are the analog of ``optical phonons'' in crystals with more than one atom per unit cell. We compute the spin wave dispersion and dynamic spin structure factor, all of which can be measured by inelastic neutron scattering experiments. We also computed ordering moment reduced by spin wave fluctuations and the uniform spin susceptibilities.
\end{abstract}
\maketitle

The discovery of the K$_x$Fe$_{2-y}$Se$_2$\cite{Kfese} system stirred up a new wave of excitement in the field of iron-based superconductors. K$_x$Fe$_{2-y}$Se$_2$ are isostructural to the ``122''-family of iron pnictides, {e.g.} BaFe$_2$As$_2$. At ambient pressure they exhibit the highest superconducting transition temperature $\approx$ 33K among iron chalcogenides\cite{Kfese}.

\begin{figure}[b]
\centering
\subfigure[]{\includegraphics[scale=1.05]{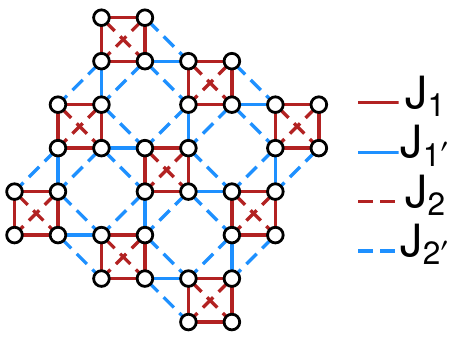}\label{fig:lattice}}
\subfigure[]{\includegraphics[scale=1.05]{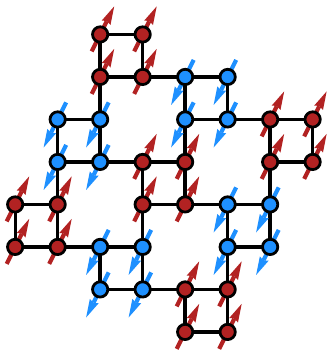}\label{fig:pattern}}
\caption{(Color online) (a) The schematic representation of the  vacancy ordered $\sqrt{5}\times\sqrt{5}$ lattice structure in each iron plane. Circles represent irons and the solid and dashed lines represent different types of spin interaction. (b) The ordered spin
configuration according to Ref.\cite{bao}. The spins point along the c-axis (out of plane).}
\end{figure}

Very recently a neutron scattering experiment\cite{bao} showed that at the
special composition $x=0.8, y=0.4$ K$_{0.8}$Fe$_{1.6}$Se$_2$ exhibits a novel
structure and magnetic state. In particular for $T<578$K the Fe vacancies
(which cause the deficiency of the iron content from 2 Fe per formula unit)
form a $\sqrt{5}\times\sqrt{5}$ superstructure in each Fe plane as shown in
Fig.\,\ref{fig:lattice}. The nearest Fe-Fe distance between the four irons
in each $\sqrt{5}\times\sqrt{5}$ unit cell is slightly smaller than that
between the neighboring cells\cite{lu,bao}.
Moreover for $T<559$K the system becomes magnetic. The
magnetic moments of the four irons in each $\sqrt{5}\times\sqrt{5}$ unit
cell align ferromagnetically along the crystalline c-axis. These aligned
``super-moments'' stagger from cell to cell to form a block-checkerboard
antiferromagnetic pattern\cite{bao}. At low temperatures the ordered moment per iron is
approximately equal to $3.3\mu_B$. Since the iron valence is found to be
+2\cite{bao3}, and LDA calculations find K$_{0.8}$Fe$_{1.6}$Se$_2$  to be an
insulator with an approximate 0.6eV energy gap\cite{dai,lu}, this implies
that each iron is in spin 2 state.

In addition to the novel structural and magnetic ordering, transport
measurement has shown samples with composition close to
K$_{0.8}$Fe$_{1.6}$Se$_2$ are superconductors with transition temperature in
the neighborhood of 30K. Ref.\cite{bao,bao3} advocates the microscopic
coexistence of superconductivity and block-antiferromagnetism. Others raises
doubt about this conclusion\cite{haihu}. We will restrain from discussing
superconductivity in this paper before the experimental situation is %completely
clear. In the following we shall focus on the magnetic property of
K$_{0.8}$Fe$_{1.6}$Se$_2$.

Given the novel magnetic order described above it is interesting to ask what
is the nature of the spin excitations.
In particular the block-antiferromagnetic order suggests in addition to the
usual gapless Goldstone modes (which can exhibit a small gap in the presence of spin-orbit interaction), there should be ``optical'' spin wave branches
analogous to the optical phonons in crystals with more than one atom per unit
cell.
Given the fact that the ordering moment is relatively large, the linearized
Holstein-Primikov Hamiltonian should be a very good approximate description
of the low energy spin dynamics.

In Ref.\cite{dai} an effective spin Hamiltonian has been used to fit the LDA results. The Hamiltonian involves intra-block nearest and second neighbor interactions $J_1,J_2$ and the inter-block nearest and second neighbor interactions $J_{1'},J_{2'}$. Because the iron magnetic moments are observed to align along the $c$-axis (perpendicular to the iron planes)\cite{bao}, there should presumably be a spin anisotropy term in the Hamiltonian. The minimum Hamiltonian we consider reads:
\begin{equation}
H=\frac{1}{2}\sum_{i,j}J_{ij}\boldsymbol{S}_i
\cdot \boldsymbol{S}_j - \Delta\sum_i S_{iz}^2,
\end{equation}
where $\Delta$ is the spin anisotropy, and $J_{ij}=J_{1}\Gamma_{ij}^{1}+J_{1'}\Gamma_{ij}^{1'}+J_{2}\Gamma_{ij}^{2}+
J_{2'}\Gamma_{ij}^{2'}$ with $\Gamma_{ij}^{b}$  ($b=1,1',2,2'$)  being the adjacency matrix defined by
\begin{equation}
    \Gamma_{ij}^{b}=
    \left\{
     \begin{array}{ll}
       1 & : (i,j) \text{~form the shortest~}b\text{-type bounds,}\\
       0 & : \text{otherwise.}
     \end{array}
   \right.
\end{equation}
The bond types are illustrated in Fig.\,\ref{fig:lattice}. The classical
ground state has the block-AFM order as shown in Fig.\,\ref{fig:pattern}. %According to Ref.\cite{bao} the ordered spins point along the crystal c-axis.
For later convenience, the sign
factor $\zeta_i=\pm1$ for $i\in$ the up(down)-spin block is defined here.

\begin{figure}[b]
\centering
\subfigure[]{\includegraphics[scale=1.0]{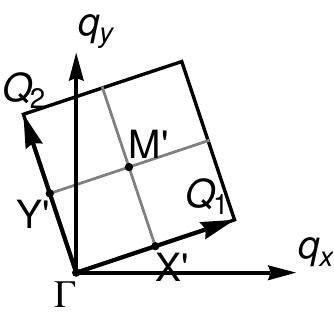}\label{fig:BZ}}
\subfigure[]{\includegraphics[scale=1.0]{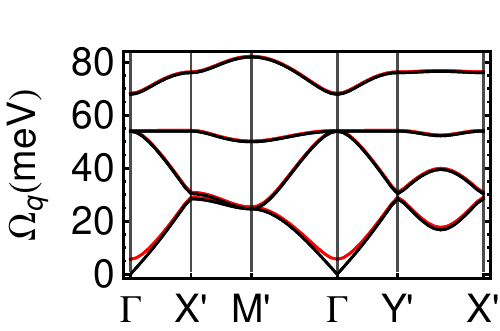}\label{fig:spectrum}}
\caption{(Color online) (a) The magnetic Brillouin zone. (b) The spin wave spectrum. The black lines represent the
dispersion in the absence of the spin anisotropy. The red lines represent the dispersion in the presence of spin anisotropy. As expected, spin anisotropy gaps the Goldstone mode.}
\end{figure}

In the following we assume each spin has $S=2$. To study the spin wave excitations
we use the linearized Holstein-Primakoff (HP) transformations $S_i^z=\zeta_i(S-a^\dag_i a_i)$ and $S^x_i=\sqrt{2S}(a^\dag_i+a_i)/2$. Using the magnetic unit cell with 8 Fe atoms, %Given the 8-Fe unit cell,
the spin Hamiltonian can be written in terms of HP bosons in the
momentum space as
\begin{eqnarray}
H=\frac{1}{2}S\left[\sum_{\boldsymbol{q}}
\phi_{\boldsymbol{q}\alpha}^\dagger
J_{\alpha\beta}(\boldsymbol{q}) \phi_{\boldsymbol{q}\beta}+O(1/S)\right],
\end{eqnarray}
where
$\phi_{\boldsymbol{q}\alpha}^\dagger
= (a_{\boldsymbol{q}\alpha}^\dagger,a_{-\boldsymbol{q}\alpha})$ and
\begin{eqnarray}
J_{\alpha\beta}(\boldsymbol{q})&=&\Big[J_{1} \Gamma_{\alpha\beta}^{1}(\boldsymbol{q})
+J_{2}\Gamma_{\alpha\beta}^{2}(\boldsymbol{q})
+J_0 \delta_{\alpha\beta}\Big]
\sigma_0\nonumber\\
&+&\Big[J_{1'}\Gamma_{\alpha\beta}^{1'}(\boldsymbol{q})
+J_{2'}\Gamma_{\alpha\beta}^{2'}(\boldsymbol{q})\Big]\sigma_1.
\end{eqnarray}
Here $\alpha,\beta$ label the 8 sites in the unit cell, $J_0=2 \Delta -2 J_{1}+J_{1'}-J_{2}+2 J_{2'}$,
and $\sigma_i$ denote the
Pauli matrices ($\sigma_0$ being the $2\times2$ identity matrix). The
Fourier transform of the adjacency matrix reads
\begin{equation}
\Gamma_{\alpha\beta}^{b}(\boldsymbol{q})
=\sum_{j, (i\in \alpha; j\in \beta)} \Gamma_{i,j}^{b} e^{i\boldsymbol{q}\cdot(\boldsymbol{r}_i-\boldsymbol{r}_j)}.
\end{equation}
Here $\boldsymbol{r}_i$ denotes the position vector of site $i$, and $\boldsymbol{q}\in$ the magnetic Brillouin zone (MBZ).
Use the the Fe-Fe bond directions as $\hat{x}$ and $\hat{y}$ and set nearest Fe-Fe bond length to unity, the MBZ is spanned by the reciprocal lattice vectors $\vect{Q}_1=(3\pi/5,\pi/5)$ and $\vect{Q}_2=(-\pi/5,3\pi/5)$. The high symmetry points are defined as $\Gamma=(0,0)$, $X'=\vect{Q}_1/2$, $Y'=\vect{Q}_2/2$, $M'=(\vect{Q}_1+\vect{Q}_2)/2$ as shown in Fig.\,\ref{fig:BZ}.

For each $\boldsymbol{q}$ we perform the Bogoliubov transformation on the
$16\times 16$ matrix $S J(\boldsymbol{q})$ to obtain the spin wave
operators and dispersions.
This can be achieved by solving the generalized eigenvalue problem
\begin{equation}
 \sum_\beta SJ_{\alpha\beta}(\boldsymbol{q})
\left(
\begin{array}{c}
 u_{n\vect{q}\beta } \\
 v_{n\vect{q}\beta }
\end{array}
\right)=
\Omega_{n\vect{q}}
%\lambda_{n\vect{q}}
\sigma_3
\left(
\begin{array}{c}
 u_{n\vect{q}\alpha } \\
 v_{n\vect{q}\alpha }
\end{array}
\right),%,
\label{eq:boson}
\end{equation}
where $\Omega_{n\vect{q}}\geq 0$ are the spin wave dispersions and   $n=1,\cdots,8$ denotes the
band index (keeping only the positive eigenvalue solutions).
Note that $\sigma_3$ is needed in Eq.(\ref{eq:boson}) for Bogoliubov transformation of bosons. The wave functions are normalized to $\sum_\alpha
(|u_{n\vect{q}\alpha}|^2 - |v_{n\vect{q}\alpha}|^2)
%\equiv \eta_{n\vect{q}} =1$.
=1$.
%Then the spin wave dispersion is given by $\Omega_{n\vect{q}}=\eta_{n\vect{q}}\lambda_{n\vect{q}}$, with the
The corresponding spin wave creation operator is given by $b^\dag_{n\vect{q}}=\sum_\alpha u_{n\vect{q}\alpha}a_{\vect{q}\alpha}^\dagger + v_{n\vect{q}\alpha}a_{-\vect{q}\alpha}$. Due to the Goldstone theorem
and the residual U(1) rotational symmetry of the spins, we anticipate two branches of degenerate gapless
spin wave modes (which would be slightly gapped by the spin anisotropy). However, since there are eight spins per magnetic unit cell, there must remain
6 gapful collective branches in the spin wave dispersion. These gapped spin wave modes are like optical phonons in the crystal with multi-atom per unit cell. These gapped spin waves are, in general, twofold degenerate due to the residual U(1) symmetry.

To explicitly demonstrate the above anticipation in the absence of the uniaxial anisotropy ($\Delta=0$) we adapt  the parameters $J_{1}=-4$meV, $J_{1'}=1$meV, $J_{2}=3$meV,
$J_{2'}=12$meV.
Aside from $J_2$ the above values are given in Ref.\cite{dai} by fitting to LDA calculations. The reason we slightly reduced $J_2$ is because using the value given in Ref.\cite{dai} we find the spin wave excitations are unstable. For the chosen parameters, the resulting spin-wave dispersion is shown as the black lines in Fig.\,\ref{fig:spectrum}. As expected, there are two degenerate Goldstone modes. The rest of the spin wave excitations form three doubly degenerate branches. In the following we shall refer to them as the``optical'' spin waves.  To demonstrate the effect of the uniaxial anisotropy we set $\Delta=0.08$meV and illustrate the corresponding spin wave dispersion using the red lines in Fig.\,\ref{fig:spectrum}.
The gap of the Goldstone mode at $\vect{q}=0$ is found to be $2S\sqrt{\Delta(\Delta +J_{1'}+2 J_{2'})}\approx6$meV. The optical modes are not much affected by the anisotropy as long
as $\Delta$ is much smaller than the magnitude of all coupling constants.

In table \ref{tab:energy}, we list the analytical expressions for  the optical spin wave excitation energies at $\Gamma$ and $M'$ (for $\Delta=0$). (Analytical expressions at other momentum points are not known to us.) In conjunction with the neutron data they can be used to determine the effective coupling constants $J_{1,2}$ and $J_{1',2'}$.

\begin{widetext}
\begin{center}
\begin{table}[t]
\caption{Table of spin wave excitation energies ($\Delta=0$)}\label{tab:energy}
\centering
\begin{tabular}{|c|c|c|}
\hline
  Momentum & Energy & Degeneracy \\ \hline
  & 0 & 2 \\
  $\Gamma$ & $2S\sqrt{(J_{1}+J_{2}-J_{2'})(J_{1}-J_{1'}+J_{2}-J_{2'})}$ & 4 \\
   & $2\sqrt{2}S\sqrt{\left(2 J_{1}-J_{1'}\right)
 \left(J_{1}-J_{2'}\right)}$ & 2 \\
 \hline
   & $2S\sqrt{J_{1}^2-(J_{1'}-2 J_{2}+2 J_{2'})
J_{1}+J_{2} (J_{2}-2 J_{2'})+J_{1'} (J_{2'}-J_{2})}$ & 4 \\
$M'$&$2S|J_{1}+\sqrt{(J_{1}-J_{2'})
(J_{1}-J_{1'}-J_{2'})}|$& 2\\
&$2S|J_{1}-\sqrt{(J_{1}-J_{2'})
(J_{1}-J_{1'}-J_{2'})}|$&2\\ \hline
\end{tabular}
\end{table}
\end{center}
\end{widetext}

\begin{figure}[b]
\centering
\subfigure[]{\includegraphics[height=0.115\textheight]{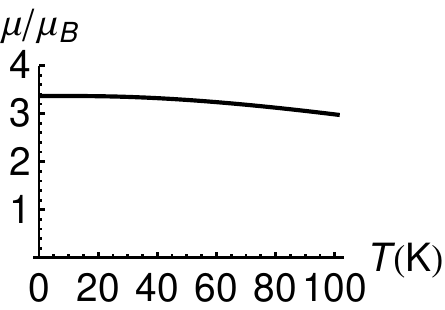}\label{fig:mu}}
\subfigure[]{\includegraphics[height=0.115\textheight]{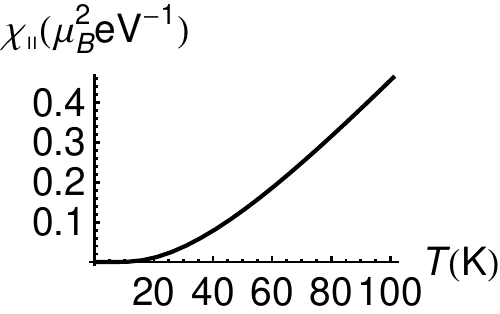}\label{fig:chizz}}
\caption{Temperature dependence of (a) the per-iron magnetic moment, and (b)
the longitudinal uniform susceptibility.}
\end{figure}

Due to quantum and thermal fluctuations, the magnetization $|\langle S_i^z\rangle|$ is reduced from $S$
to $|\langle S_i^z\rangle| =S - \delta S$, where the correction $\delta S$ is given by
\begin{equation}
\begin{split}
\delta S &=
\int_{\vect q\in \textrm{MBZ}}\frac{d^2\vect{q}}{(2\pi/\sqrt5)^2}
%\sum_{\vect{q}\in\text{MBZ}}
\langle a_{\vect{q}\alpha}^\dagger
a_{\vect{q}\alpha}\rangle,\\
%&= \sum_{n,\vect{q}\in\text{MBZ}}
&=\int_{\vect q\in\textrm{MBZ}}\frac{d^2\vect{q}}{(2\pi/\sqrt5)^2}
\sum_n\Big[|u_{n\vect{q}\alpha}|^2n_B(\Omega_{n\vect{q}})\\ &~~~~~~~~~~~~~~~~~~~~~~ +|v_{n\vect{q}\alpha}|^2(n_B(\Omega_{n\vect{q}})+1)\Big],
\end{split}
\end{equation}
where $n_B$ is the Bose-Einstein distribution function and there is no summation over $\alpha$. The moment correction
$\delta S$ is found to be the same on all sites. For the parameter
 $J_{1}=-4, J_{1'}=1, J_{2}=3, J_{2'}=12, \Delta=0.08$meV, we find $\delta S =0.32$ at zero temperature.  This amounts to a magnetic moment per iron equal to
$2\times(S-\delta S)=3.36$ Bohr magneton ($\mu_B$), in good agreement with the experimental deduced value. The
temperature dependence of the ordering moment (per iron) is shown in Fig.\,\ref{fig:mu}.

\begin{figure}[b]
\centering
\includegraphics[scale=1.0]{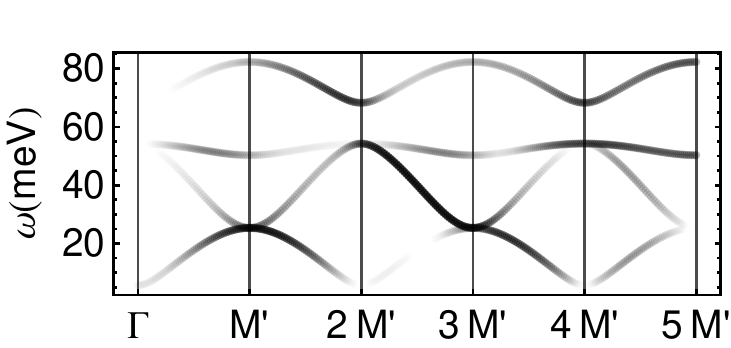}
\caption{Dynamic spectral function along the cut through the
$\Gamma$-$M'$ line. $pM'$ denotes the momentum point of
$(p\pi/5,2p\pi/5)$. $2M'$ and $4M'$ are equivalent to $\Gamma$ point
in the MBZ, while $3M'$ and $5M'$ are equivalent to
$M'$. The spin wave dispersion is plotted using the spectral weight of the dynamic spectral function as darkness. Darker line implies greater spectral weight.}\label{fig:weight}
\end{figure}

To the same leading order approximation, the zero-temperature dynamic %spectral function
structure factor is given by
\begin{equation}
  S_\perp(\vect{q},\omega)=
  \sum_{n}A_{n\vect{q}}
  %\frac{1-\eta_{n\vect{q}}}{2}
  \delta(\omega-\Omega_{n\vect{q}}),
\end{equation}
where the spectral weight $A_{n\vect q}$ reads
\begin{equation}
A_{n\vect{q}}=\sum_{\alpha,\beta}
(u_{n\vect{q}\alpha}^*+v_{n\vect{q}\alpha}^*)
(u_{n\vect{q}\beta}+v_{n\vect{q}\beta})
e^{i\vect{q}\cdot(\vect{r}_\alpha-\vect{r}_\beta)}.
\end{equation}
Note that $A_{n\vect q}$ is not periodic with reciprocal magnetic unit vector $\vect Q_1$ and $\vect Q_2$ even though spin wave energy $\Omega_{n\vect q}$ is. This is similar to the matrix element effect in angle-resolved photoemission experiments.
Fig.\,\ref{fig:weight} shows the dynamic structure factor along the $\Gamma$-$M'$ cut. In inelastic neutron scattering experiments, the $\Gamma$-point ``optical'' modes will be observable at $(2\pi/5,4\pi/5)$ point, and the $M'$-point ``optical'' modes will be visible at $(\pi,0)$ point.

It is also straightforward to compute the uniform susceptibility. The transverse uniform susceptibility reads
\begin{equation}
\chi_\perp=\frac{S \mu_B^2}{128}\lim_{\vect{q}\rightarrow 0}
\sum_{n,\alpha,\beta}
\frac{(u_{n\vect{q}\alpha}^*+v_{n\vect{q}\alpha}^*)
(u_{n\vect{q}\beta}+v_{n\vect{q}\beta})}
{\Omega_{n\vect{q}}},
\end{equation}
which is independent of temperature in the low temperature regime. Using our parameters, it can be evaluated to $\chi_\perp=2.49\mu_B^2\text{eV}^{-1}$.
The longitudinal uniform susceptibility is given by
\begin{eqnarray}
%\chi_\parallel= \frac{\mu_B^2}{256}\sum_{n,\vect{q},\alpha}\tilde{\zeta}_{n\vect{q}}^2n_B^\prime(\Omega_{n\vect{q}}),\\
\chi_\parallel= \frac{\mu_B^2}{256}\sum_{n,\vect{q}} \left[\sum_\alpha\zeta_\alpha(|u_{n\vect{q}\alpha}|^2
+ |v_{n\vect{q}\alpha}|^2)\right]^2
n_B^\prime(\Omega_{n\vect{q}}),
\end{eqnarray}
where %$\tilde{\zeta}_{n\vect{q}}
%=\sum_\alpha\zeta_\alpha(|u_{n\vect{q}\alpha}|^2+ |v_{n\vect{q}\alpha}|^2)$.
where %$\alpha$ labels the four-iron block,
$\zeta_\alpha=\pm 1$ for $\alpha$ residing in spin up or down block respectively, and $n_B^\prime$ denotes the  derivative of
Bose-Einstein distribution function. Fig.\,\ref{fig:chizz} shows
the temperature dependence of $\chi_\parallel$. Below the
temperature scale of the spin gap ($\sim$50K), $\chi_\parallel$
exhibit exponential temperature dependence. While above that
temperature scale, the linear-$T$ behavior $\chi_\parallel
\propto T$ can be observed.

In conclusion we have studied the spin wave excitations of the block-antiferromagnetic K$_{0.8}$Fe$_{1.6}$Se$_2$ using an effective spin model suggested by LDA calculation. There are both the Goldstone modes and the ``optical'' spin wave branches. After the spin wave correction, the ordering moment is found to be $\approx 3.4\mu_B$ per iron at low temperatures, which is close to the observed value. The observation of the optical spin wave modes will constitute a spectacular supporting evidence for the block antiferromagnetism.

Currently we do not have a good understanding of the block-antiferromagnetism and its relation to the structural distortion from microscopic point of view. Understanding it can shed light on the local electronic structure and correlation for the newly discovered K$_x$Fe$_{2-y}$Se$_2$ and other pnictides.

\begin{acknowledgments}
{\it Acknowledgement}: This work is supported, in part, by DOE grant number DE-AC02-05CH11231 (DHL and HY), and by China Scholarship Council and Tsinghua Education Foundation in North America (YZY).
\end{acknowledgments}

\end{document}